\documentclass{mnras}

\usepackage[hyphenbreaks]{breakurl}

\title[{\sl Swift observations of the symbiotic binary AP Peg} ]{ {\sl
    Swift} observations of the 2015 outburst of AG Peg -- from
  slow nova to classical symbiotic outburst}

\author[]
{Gavin Ramsay$^{1}$, J. L. Sokoloski$^{2}$, G. J. M. Luna$^{3}$,  N. E. Nu\~{n}ez$^{4}$\\
$^{1}$Armagh Observatory, College Hill, Armagh, BT61 9DG, UK\\
$^{2}$Columbia Astrophysics Lab, 550 W120th St., 1027 Pupin Hall, MC 5247 Columbia University, 10027, New York, USA\\
$^{3}$Instituto de Astronom\'ia y F\'isica del Espacio (IAFE, CONICET-UBA), Av. Inte. G\"uiraldes 2620, C1428ZAA, Buenos Aires, Argentina\\
$^{4}$Instituto de Ciencias Astron\'{o}micas de la Tierra y el Espacio (ICATE-UNSJ), Av. Espan\~{a} (sur) 1512, 5400, 
San Juan, Argentina\\
}

\date{Accepted 2016 June 23. Received 2016 June 23; in original form 2016 April 5}

\begin{document}
\outer\def\gtae {$\buildrel {\lower3pt\hbox{$>$}} \over 
{\lower2pt\hbox{$\sim$}} $}
\outer\def\ltae {$\buildrel {\lower3pt\hbox{$<$}} \over 
{\lower2pt\hbox{$\sim$}} $}
\newcommand{\Msun} {$M_{\odot}$}
\newcommand{\lsun} {$L_{\odot}$}
\newcommand{\Rsun} {$R_{\odot}$}
\newcommand{\solar} {${\odot}$}
\newcommand{\kep}{\it Kepler}
\newcommand{\swift}{\it Swift}
\newcommand{\Porb}{P_{\rm orb}}
\newcommand{\nuorb}{\nu_{\rm orb}}
\newcommand{\eplus}{\epsilon_+}
\newcommand{\eminus}{\epsilon_-}
\newcommand{\cd}{{\rm\ c\ d^{-1}}}
\newcommand{\MdotL}{\dot M_{\rm L1}}
\newcommand{\Mdot}{$\dot M$}
\newcommand{\Mdotsolar}{\dot{M_{\odot}} yr$^{-1}$}
\newcommand{\Ldisk}{L_{\rm disk}}
\newcommand{\src}{KIC 9202990}
\newcommand{\ergscm} {erg s$^{-1}$ cm$^{-2}$}
\newcommand{\rchi}{$\chi^{2}_{\nu}$}
\newcommand{\chisq}{$\chi^{2}$}
\newcommand{\pcmsq} {cm$^{-2}$}

\maketitle
\begin{abstract}
Symbiotic stars often contain white dwarfs with quasi-steady shell
burning on their surfaces. However, in most symbiotics, the origin of
this burning is unclear.  In symbiotic slow novae, however, it is
linked to a past thermonuclear runaway.  In June 2015, the symbiotic
slow nova AG Peg was seen in only its second optical outburst since
1850.  This recent outburst was of much shorter duration and lower
amplitude than the earlier eruption, and it contained multiple peaks
-- like outbursts in classical symbiotic stars such as Z And.  We
report Swift X-ray and UV observations of AG Peg made between June
2015 and January 2016.  The X-ray flux was markedly variable on a time
scale of days, particularly during four days near optical maximum,
when the X-rays became bright and soft. This strong X-ray variability
continued for another month, after which the X-rays hardened as the
optical flux declined.  The UV flux was high throughout the outburst,
consistent with quasi-steady shell burning on the white dwarf.  Given
that accretion disks around white dwarfs with shell burning do not
generally produce detectable X-rays (due to Compton-cooling of the
boundary layer), the X-rays probably originated via shocks in the
ejecta.  As the X-ray photo-electric absorption did not vary
significantly, the X-ray variability may directly link to the
properties of the shocked material.  AG Peg's transition from a slow
symbiotic nova (which drove the 1850 outburst) to a classical
symbiotic star suggests that shell burning in at least some symbiotic
stars is residual burning from prior novae.
\end{abstract}


\section{Introduction}

Symbiotic Binaries have orbital periods with timescales of years and
contain a cool giant star which is losing mass, typically via a wind,
to a hot compact star (see Allen 1984, Kenyon \& Webbink 1984 and
Mikolajewska 2007). The evolution of these systems has long been
debated (e.g. Iben \& Tutukov 1996), but there is now a growing
appreciation that symbiotic binaries could produce some fraction of
supernovae Ia outbursts (e.g. Di Stefano 2010, Dilday et al. 2012).

In their catalogue of symbiotic stars, Belczy\'{n}ski et al. (2000)
list more than 200 known or suspected symbiotic stars, with more being
discovered through surveys such as IPHAS (Corradi et al. 2008, 2010,
Rodr\'{i}guez-Flores et al. 2014). The optical photometric
characteristics of symbiotic stars are diverse. Symbiotic stars can
show `classical' outbursts lasting weeks to years where they brighten
by $\sim$1--2 mag. It is unclear if most of these outbursts are due to
nuclear burning on the surface of the white dwarf or an accretion disk
instability of the sort seen in some cataclysmic variables (CVs) such
as dwarf novae or a combination of both nuclear burning and a disk
instability (as in Z And, Sokoloski et al. 2006). There are other
symbiotic stars which show outbursts which are powered by
thermonuclear runaways (TNRs), including slow novae where the outburst
event can extend for many decades (e.g. Allen 1980, Kenyon \& Truran
1983).

Observations made using the {\sl Einstein} and {\sl ROSAT} satellites
showed that symbiotic stars are X-ray emitters at relatively low flux
levels when in a quiescent state (Allen 1981, M\"{u}rset, Wolff \&
Jordan 1997). Luna et al. (2013) made a survey of symbiotic stars
using the X-ray telescope (XRT) on-board the {\swift} satellite and
found (as did M\"{u}rset et al. 1997) that they could group the
sources into those which had very soft X-ray spectra (originating from
shell burning on the surface of the white dwarf); those which had
emission extending to $\sim$2.4 keV (probably due to the collision of
winds from the red giant and white dwarf); and those which showed
emission above $\sim$2.4 keV. Whereas M\"{u}rset et al. (1997) ascribed
the hard X-rays to accretion onto a neutron star, Luna et al. (2013)
attributed the thermal hard X-ray emission, as detected using the {\sl
  Swift} XRT, to boundary layer emission around a white
dwarf. Inevitably some sources showed X-ray spectra combining
characteristics of more than one of these classes. These observations
were generally made when the symbiotic binary had been in quiescence
and not in outburst.

Most X-ray observations of symbiotic stars have typically been made at
only one epoch. In contrast, observations of the symbiotic nova AG Dra
made using {\sl ROSAT} in 1994 show that the X-ray flux
decreased over the course of the optical outburst (which was also
noted in an earlier minor outburst using {\sl EXOSAT} observations)
which was attributed to the decrease in the temperature of the hot
component, possibly due to an increase in the apparent size of its
outer envelope (Greiner et al. 1997). An anti-correlation between the
optical and X-ray flux over an outburst has also been seen in dwarf
novae (e.g. Wheatley et al. 1996) and in the double degenerate
interacting helium binary KL Dra (Ramsay et al. 2012).

The symbiotic binary AG Peg is one example of a slow symbiotic nova,
which has a photometric record extending more than 200 years. At the
beginning of the 19th century AG Peg was recorded as a $\sim$9.5 mag
star which experienced a gradual brightening around 1850, reaching
$\sim$6 mag 20 years later (Rigollet 1947, Belyakina 1968).
(M\"{u}rset \& Nussbaumer 1994 show the light curve of AG Peg
extending over 150 years.) Observations made by amateur astronomers
whose records are accessible on-line
(e.g. AAVSO\footnote{\url{http://aavso.org}} and
BAAVSS\footnote{\url{http://britastro.org/vssdb}}) date back to
1941. These show that AG Peg made a slow decline in brightness and
only reached its pre 1850 outburst by the end of the 20th
century. Indeed, there is evidence that the 1850 outburst was the
slowest nova outburst ever recorded (e.g. Kenyon, Proga \& Keyes
2001).

As AG Peg is a relatively bright object, it was studied
spectroscopically by the end of the 19th century, showing a range of
emission and absorption lines whose characteristics changed over months
and decades. Later observations show the binary orbital period is
818.2$\pm$1.6 d with an eccentricity of 0.11 (Merrill 1951, Fekel et
al. 2000).  The spectral type of the giant star is M3 III (Kenyon \&
Fernandez-Castro 1987), while the mass function of the system is
consistent with the hot component being a white dwarf (Fekel et
al. 2000).

There was some indication in May 2013 that AG Peg was beginning a new
active phase when it was reported that AG Peg had brightend by
$\sim$0.3 mag (Munari et al. 2013). However, during June 2015 it was
found that AG Peg had brightened from $V\sim$8.5 to $V\sim$7.0 over
the course of a few weeks (Waagen 2015). Based on the long term
  light curve shown in M\"{u}rset \& Nussbaumer (1994) and the records
  of the amateur astronomer groups mentioned earlier, we consider it
  highly unlikely that an outburst of the duration and amplitude of
  the 2015 outburst has taken place on AG Peg since the early 1940's,
  and the 2015 outburst is therefore likely to be the first {\sl bona
    fide} outburst since the 19th century. However, we cannot exclude
  the possibility that a short duration event took place, for instance
  during the intervals when AG Peg was close to the Sun and therefore
  not observable.

We requested Target of Opportunity observations using {\sl Swift}
shortly after the announcement of the 2015 eruption and these
observations showed that AG Peg was detected in X-rays during the
optical outburst (Luna et al. 2015, Ramsay et al. 2015). Since these
early observations, {\swift} has been used to observe AG Peg at
regular intervals. The main goal has been to determine how the X-ray
flux (and spectrum) changes over the outburst and compare it with the
observations of other accreting white dwarfs.

\begin{figure}
\begin{center}
\setlength{\unitlength}{1cm}
\begin{picture}(6,9)
\put(-1.,-1.8){\includegraphics{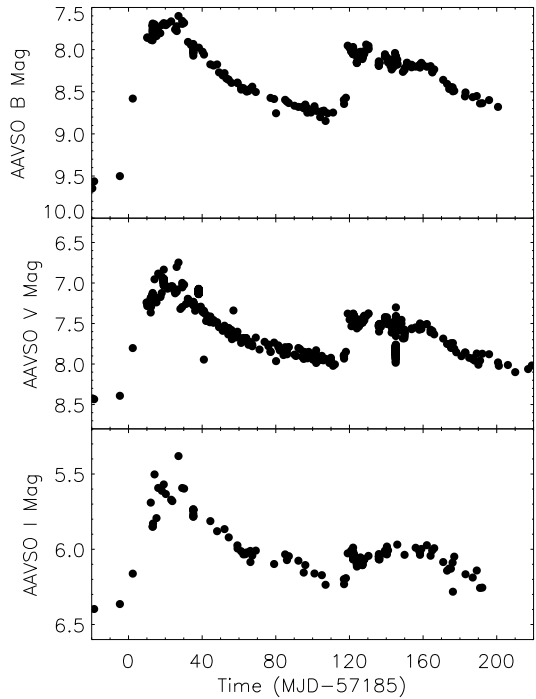}}
\end{picture}
\end{center}
\caption{Photometry of AG Peg over the outburst in the $BVI$ bands and
  taken from AAVSO data (Kafka 2015). We have not attempted to remove
  potentially discrepant data points. The outburst started on
  MJD=57185 which corresponds to 2015 June 12.}
\label{aavso-bands} 
\end{figure}

\section{Overview of observations made by amateur observers}

\subsection{Photometry}

Observations made by amateur astronomers (e.g. AAVSO and BAA VSS) show
that AG Peg was at $V\sim$8.5 mag on 2015 June 6, reaching $V\sim$7.5
by June 21 and reached a peak of $V\sim$6.8 mag by the end of June
(Figure \ref{aavso-bands}). Over the next 40 days it declined by half
a magnitude after which it declined more slowly, reaching a plateau
phase at $V\sim$7.8 mag. Around Oct 10 AG Peg showed a
`re-brightening' event reaching $V\sim$7.5 mag after$\sim$20 days,
after which it started to decline. The peak magnitude of the June
outburst was $\sim$1 mag fainter than it was in 1870 and the rate of
decline in this current outburst appears much quicker (e.g. Belyakina
1968). The amplitude of the outburst is higher in the $B$ band
($\sim$1.8 mag) compared to the $I$ band ($\sim$0.9 mag). These data
also show the re-brightening event ($V\sim$0.5 mag) occured on a
timescale of less than 1 d and was relatively brighter at bluer
wavelengths compared to the June 2015 outburst.

\subsection{Spectroscopy}
\label{opt-spec}

AG Peg was observed using the LOTUS NUV-optical spectrograph on the
Liverpool Telescope on 2015 July 1 (approximately 20 days after of the
start of the outburst). Compared to pre-outburst spectra, the relative
strength of O and N emission lines relative to the Balmer lines was
stronger during the outburst (Steele et al. 2015).

Many groups, including amateur
astronomers\footnote{e.g. \url{http://www.astronomie-amateur.fr}},
have made spectroscopic observations of AG Peg over its outburst.
These observations show strong emission lines, including the Balmer
series plus He {\sc i} (6678) and He {\sc ii} (4686), [O {\sc iii}]
(4363) and the O {\sc vi} emission band at 6825 \AA, which is due to
Raman scattering. A full analysis of the optical spectra made over the
course of the outburst is beyond the scope of this work, but we were
able to estimate the effective temperature of the ionizing
source using the He {\sc ii} (4686) and H$\beta$ lines and the formula
of Iijima (1981) (quoted in Sokoloski et al. 2006) which derives the
effective temperature using the equivalent width (EW) of these
lines (we ignore the He {\sc i} (4471) line since it is much weaker
than He {\sc ii} and H$\beta$).

\begin{figure}
\begin{center}
\setlength{\unitlength}{1cm}
\begin{picture}(6,10)
\put(-1,-0.8){\includegraphics{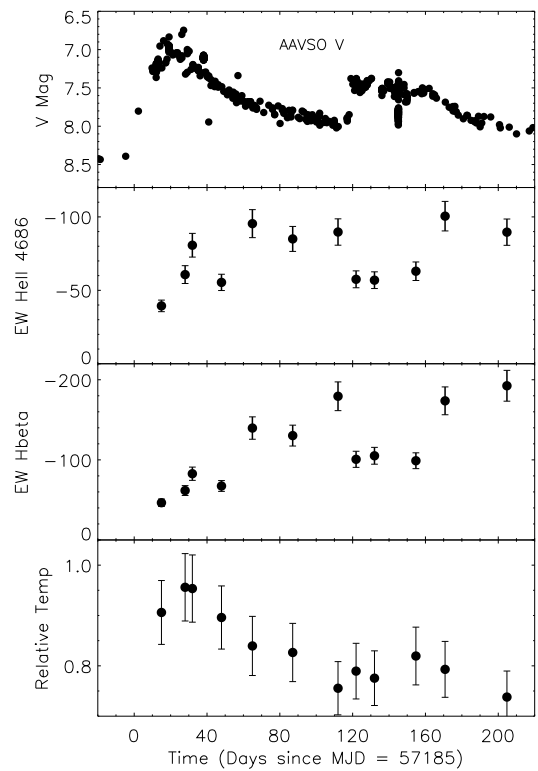}}
\end{picture}
\end{center}
\caption{From the top we show the AAVSO V band light curve; the
  equivalent width of the He {\sc ii} (4686) and H$\beta$ line
  (derived using data taken by French amateur astronomers) and the
  relative effective temperature derived using these lines.
    As there is some uncertainty over the absolute temperature
    calibration using this method, we determine the relative
    temperature over the course of the outburst, where a value of
    unity implies an effective temperature of 1.9$\times10^{5}$ K
    using the formula of Iijima (1981).}
\label{opttemp} 
\end{figure}

We used spectra taken by amateur
astronomers\footnote{\url{http://www.astrosurf.com/aras/Aras\_DataBase/Symbiotics/AGPeg.htm}}
which covered the He {\sc ii} (4686) and H$\beta$ lines to derive
their EW (we estimated that the error on the EW measurements was
$\sim$10 percent by making a number of measurements of the same
line). This shows that the EW of He {\sc ii} (4686) was somewhat
variable over the outburst but showed a distinct decrease at the time
of the optical re-brightening event (Figure \ref{opttemp}). This
decrease is also seen in the EW of the H$\beta$ line, although there
is a trend for the EW to increase over the course of the outburst.  We
show the relative temperature over the course of the outburst in
Figure \ref{opttemp} which shows that the effective temperature
of the ionizing source decreases over the duration of the outburst.

\section{{\sl Swift} observations}

\begin{figure*}
\begin{center}
\setlength{\unitlength}{1cm}
\begin{picture}(6,10)
\put(12,-0.8){\includegraphics{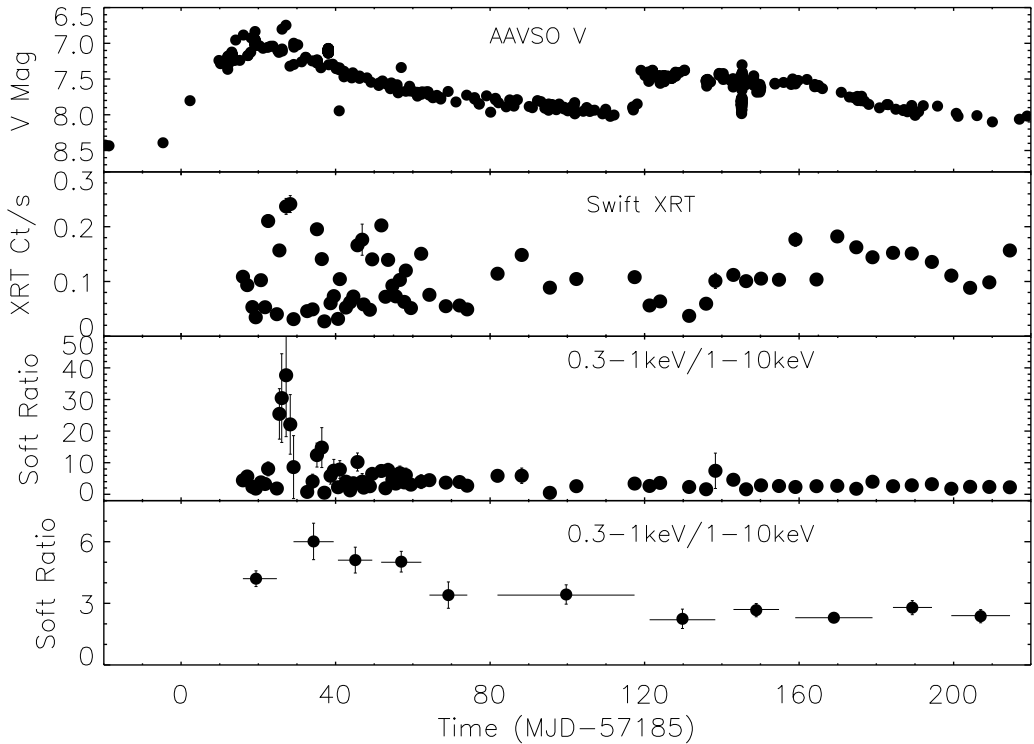}}
\end{picture}
\end{center}
\caption{From the top we show: The $V$ magnitude of AG Peg as obtained
  by AAVSO observers; the X-ray light curve of AG Peg (0.3--10 keV) as
  derived using {\sl Swift} XRT data; the X-ray softness ratio; and
  the softness ratio where we have combined a number of epochs of data
  and (for clarity) we have omitted the epoch with the softest data
  points.}
\label{light} 
\end{figure*}

\subsection{Overview}

The {\swift} satellite was launched in 2004 with a primary goal of
detecting gamma-ray bursts, and locating and characterising their
X-ray and optical/UV properties (Gehrels et al. 2004). NASA runs a
Target of Opportunity programme which provides the means for observing
targets which have unexpectedly become interesting (novae for
instance) at the earliest opportunity. Observations of AG Peg were
made using {\swift} on 2015 June 28th. Given the brightness of the
source, it was saturated in the UVOT instrument (we discuss the
implications of this in \S 4.3).

Because {\swift} is in a low Earth orbit, each observation sequence
(which makes up an `ObservationID') can be made up of more than one
pointing, while the total exposure time of the X-ray observation in
each ObservationID was typically 1--2 ksec (see Table \ref{xrtlog}).
The XRT (Burrows et al. 2005) on-board {\swift} has a field of view of
23.6$\times$23.6 arcmin with CCD detectors allowing spectral
information of X-ray sources to be determined. It is sensitive
  over the range 0.3--10 keV and at launch had a spectral resolution
  of 140 eV at 6 keV (which has since degraded over time).

The XRT has a number of modes of operation (designed to observe
gamma-ray bursts at various stages of their evolution), but here we
concentrate on the `photon counting' mode which has full imaging and
spectroscopic information. The data are processed using the standard
XRT pipeline and it is these higher level products which we use in our
analysis {\footnote{\url{http://swift.gsfc.nasa.gov/docs/swift/analysis/
xrt\_swguide\_v1\_2.pdf}}.

\subsection{X-ray light curve}

To determine the count rate of AG Peg at each epoch we filtered the
X-ray events to extract photons with corresponding `grades' 0--12
(this has the effect of filtering out events which are not from the
target) and energy ranges from each ObservationID and made an image
using {\tt
  xselect}\footnote{\url{http://heasarc.nasa.gov/docs/software/lheasoft/ftools/xselect/xselect.html}}. We
used the HEASoft tool {\tt XIMAGE} and the routine {\tt SOSTA} (which
takes into account effects such as vignetting, exposure and the point
spread function) to determine the count rate and error at the position
AG Peg. We determined an energy independent correction for each
  pointing to account for the fact that X-ray photons from AG Peg may
  have been deposited on known bad columns on the XRT detectors using
  the tool {\tt
    xrtlccorr}\footnote{\url{http://www.swift.ac.uk/analysis/xrt/lccorr.php}}.}
We show the count rate as a function of time since the start of
  the outburst in Figure \ref{light}. (The characteristics of
  the X-ray light curve are very similar whether we apply the
  correction or not).

The first four epochs of the X-ray observations were made as AG Peg
was just approaching maximum optical brightness and show a decreasing
X-ray flux ($\sim$0.07 to 0.01 ct/s). This compares with a count rate
of 0.02--0.03 ct/s when AG Peg was observed in 2013 Aug using the XRT.
The X-ray flux then experiences some epochs of significantly enhanced
X-ray rates reaching $\sim$0.3 ct/s at MJD=57211 (26 d after the start
of the outburst). As AG Peg slowly declines in optical flux the X-ray
flux continues to show significant variations, although there is a
decline in the peak count rate over this time interval.

As indicated earlier, approximately 120 days after the start of
  the optical outburst there is an optical rebrightening event where
the system increases in brightness by $\sim$0.5 mag. In X-rays there
is a decline in flux as the system brightens in the optical. However,
by the time the system reaches maximum optical brightess and starts to
decline, the X-ray flux starts to increase in flux, giving tentative
evidence for a delay of $\sim$20 days between the X-ray and optical
flux, or a supression of X-rays during the rebrightening episode
(Figure \ref{light}).

To search for changes in the broad X-ray spectral distribution over
the outburst we derived light curves (using the same method as before)
for two energy bands: 0.3--1.0 keV and 1--10 keV (for reference we
give the count rate in different energy bands in Table
\ref{xrtlog}). We show how the softness ratio (0.3--1.0 keV/1--10 keV)
changes over the outburst in Figure \ref{light}. It is clear that
during the time of highest flux (MJD$\sim$57210) the softness ratio is
high (i.e. the spectrum is dominated by soft X-ray photons). As the
source declines in optical flux, the large day-to-day X-ray flux
variations are not reflected in changes to the softness ratio whose
changes are much less marked. We also co-added data from different
epochs to search for a variation in the softness ratio as the outburst
progressed and the results of this are shown in Figure
\ref{light}. The spectrum of AG Peg appears to harden after the very
soft epoch reaching a minimum by the time of the onset of the
rebrightening event. During the rebrightening event the softness ratio
is roughly constant.

\subsection{X-ray spectra}

We now investigate the X-ray spectrum of AG Peg in more
detail. Photons were extracted from a circular aperture centered on AG
Peg and also a source-free region of the detector to create a
backround spectrum. We obtain an ancillary file, which allows us to
determine the flux of the source, using an exposure map derived from
the event files which are used to make up the spectrum. We used the
response file, which allows the energy of each photon to be tagged,
appropriate to the mode and grade of event, which was taken from the
NASA HEASARC
site\footnote{\url{http://swift.gsfc.nasa.gov/proposals/swift\_responses.html}}. The
source spectrum was binned so that each bin had a minimum of 20 counts.

Using all the data, with the exception of the soft bright state, we
obtain a spectrum with a total exposure time of 113.3 ksec. This
spectrum is consistent with that of absorbed thermal plasma emission
with emission detected up to $\sim$3 keV. M\"{u}rset et al. (1997) and
Luna et al. (2013) classed the X-ray properties of Symbiotic binaries
according to the nature and energetics of the spectrum. The integrated
spectrum of AG Peg during outburst is similar, for instance, to
{\swift} J1719--3002 made during a quiescent state (Luna et al. 2013)
which was classed as having a '$\beta$' spectrum which indicates a
spectrum peaking $\sim$0.8 keV, with most photons softer than 2.4 keV
and is likely produced by a collision of winds from the white dwarf
and red giant ($\sim$1/3 of sources in the study of Luna et al. show a
$\beta$ type spectrum).

\begin{figure*}
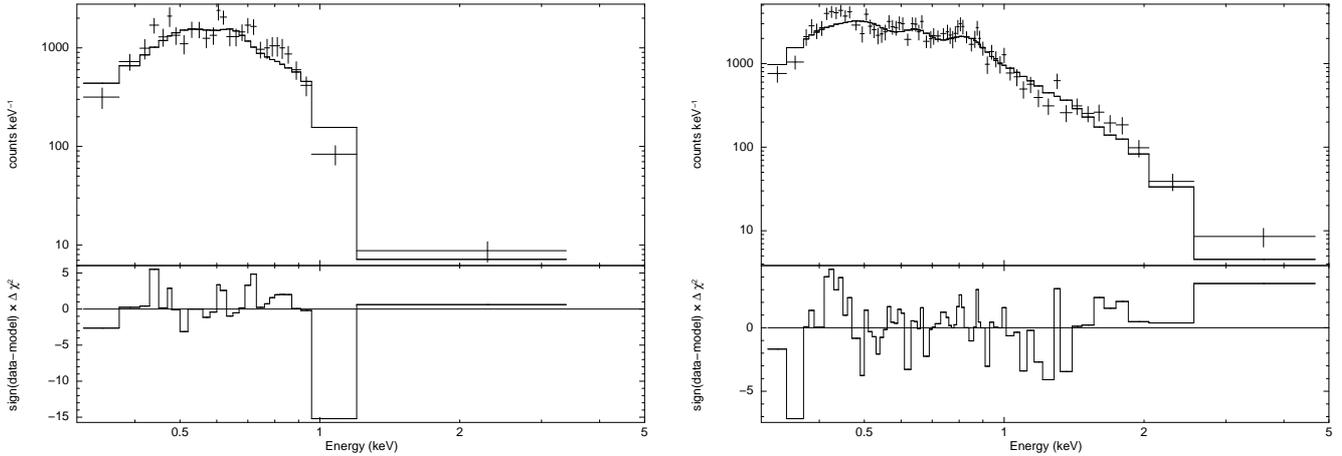

\begin{center}
\setlength{\unitlength}{1cm}
\begin{picture}(10,6)
\put(5,6.8){\includegraphics{outspec-src.ps}}
\put(-4,6.8){\includegraphics{soft-newspec.ps}}
\end{picture}
\end{center}
\caption{On the left hand side we show the X-ray spectrum taken during
  the soft X-ray phase and on the right the X-ray spectrum taken
  immediately after this (the `high' phase). The model best fit is
  shown as a solid line and is an two temperature absorbed thermal
  plasma model.}
\label{xrtspec} 
\end{figure*}

Using the X-ray fitting package {\tt XSPEC} (Dorman \& Arnaud
  2001), we modelled the integrated spectrum using a range of emission
  models in combination with the {\tt tbabs} absorption model (Wilms
  et al. 2000). We found that although a two temperature thermal
  plasma model gave significantly better fits than a one temperature
  model it was not formally a good fit (\rchi=2.60). We allowed
  various elements to have non-Solar abundances and whilst we obtained
  formally better fits, it was not a formally acceptable fit
  (\rchi=1.59). In addition, there was some concern that we were
  obtaining non-physical abundances and therefore fitting noise in the
  spectrum.

The observed flux inferred from this model is $1.3\times10^{-12}$
\ergscm (0.3--10 keV); the unabsorbed flux in the same band is
1.9$\times10^{-12}$ \ergscm, giving an X-ray luminosity, assuming a
distance of 1 kpc (interferometric observations show AG Peg to have a
parallax less than 1 mas, Boffin et al. 2014) of
$L_{X}\sim2.3\times10^{32}$ erg/s (which is typical for symbiotic
stars in a quiescent state, c.f. Luna et al. 2013). 

It is clear from the softness ratio shown in Figure \ref{light} that
the X-ray spectrum is much softer $\sim$25 days after the start of the
outburst. We extracted a spectrum from this soft phase and for
comparison, we extracted a spectrum from data taken from pointings
made in the next 30 days after this (the `high' phase). We show the
X-ray spectrum taken from the soft and high phases in Figure
\ref{xrtspec}. There is no evidence for a soft blackbody component in
the `soft' spectrum -- rather the soft spectrum was best fit by a
single thermal plasma model whilst the high spectrum was best fit
using two thermal plasma models. The difference between the 'soft' and
'high' spectra was not due to different amounts of
absorption. However, the goodness of fits to both spectra were still
not formally acceptable (\rchi=2.28 and \rchi=1.63 for the soft and
high spectra respectively). We therefore do not provide additional
details of the fits.

\section{Discussion}

The 2015 outburst of AG Peg is the first to be observed since the mid
19th century. The duration and amplitude of these outbursts are very
different in the optical, with the most recent being much shorter and
of lower amplitude compared to the earlier outburst.  We now discuss
the physical reasons for these differences, the origin of the X-ray
emission, and the generation of quasi-steady nuclear burning on the
white dwarfs in AG Peg and other symbiotic binaries.

\subsection{The nature of the 2015 outburst}

The `slow-nova' outburst of AG Peg in the mid 19th century was almost
certainly driven by a TNR on the surface of the accreting white dwarf
(e.g. Gallagher et al. 1979, Kenyon et al. 1993). In contrast, the
double-peaked 2015 outburst was much less energetic and of shorter
duration. Moreover, whereas optical spectra taken during the decline
of the slow-nova showed only absorption lines due to the expanded
white dwarf photosphere (Gallagher et al. 1979), optical spectra
during the 2015
eruption\footnote{e.g. \url{http://www.astrosurf.com/aras/novae/InformationLetter/ARAS\_EruptiveStars\_2016-01.pdf}}
displayed emission lines such as O {\sc vi} (indicating highly ionized
species in the nebula) which imply that material was still being
exposed to the photoionizing hot white dwarf.

Rather than mimicking a slow-nova, the 2015 outburst appeared more
similar to {\sl classical} symbiotic outbursts which have time scales
of months to years (for instance Z And or AG Dra).  The coverage of
the optical photometry during the rise to optical maximum was not high
enough to reveal whether the eruption was triggered by an
accretion-disk instability, as in Z~And. However, the optical light
curve of Z And during the 2000--2002 event (shown in Sokoloski et
al. 2006), also shows a rebrightening event and is remarkably like AG
Peg, the difference being that in Z And the outburst lasted 2 years
rather than 6 months in the case of AG Peg. The re-brightening event
is also reminiscent of brightness oscillations seen in classical
symbiotic outbursts in e.g. CI~Cyg, AX~Per, and AG~Dra (Mikolajewska
\& Kenyon 1992; Viotti et al. 2005), which some authors have
speculated could be driven by resonances in an accretion disk. The
modest increase in X-ray flux averaged over the 2015 event is also
consistent with the increases in X-ray flux from Z And during its
classical symbiotic outburst.

\subsection{The origin of the X-ray emission}

It is thought that classical outbursts from symbiotics may be driven
by a disk instability, of the sort which drives dwarf nova outbursts,
which can increase the rate of quasi-steady nuclear shell burning on
the white dwarf (Sokoloski et al. 2006).  However, whereas almost all
dwarf nova outbursts in CVs show a suppression of X-rays as the
boundary layer between the accretion disk and the white dwarf becomes
optically thick (with temperatures of just a few tens of eV; Wheatley
et al. 1996), the connection between X-ray emission and the underlying
physics is less clear in most classical symbiotic outbursts.  In
symbiotic stars with quasi-steady shell burning on the white dwarf,
the accretion disk does not produce detectable X-ray emission because
UV photons from the hot white dwarf Compton cool the boundary layer
out of the X-ray regime (e.g., Luna et al. 2013).  As discussed in \S
\ref{shellburning}, there is evidence that the white dwarf in AG Peg
hosted shell burning throughout the 2015 eruption.

The temperature of the X-ray emission in AG Peg is much lower than the
temperatures typically seen in dwarf novae in quiescence ($\sim$10
keV) -- consistent with the idea that the X-ray emitting plasma in AG
Peg was not located near the surface of the white dwarf, deep within its
potential well.  Our spectral models indicate temperatures below 1
keV.  Super-Soft X-ray emission is commonly seen after outbursts
from novae and occasionally from symbiotics (e.g. AG Dra, Skopal et
al. 2009). However, as expected for a low-mass white dwarf with
quasi-steady shell burning, no super-soft X-ray emission was detected
in AG Peg, confirming that nuclear burning did not lead to a white
dwarf photosphere which was hot enough for blackbody emission to be
seen in the supersoft X-rays.

One of the characteristics of the X-ray observations of AG Peg is the
high day-to-day variability. X-ray observations of AG Peg started with
{\swift} shortly before it reached peak optical brightness. They show
that around 20--30 days after the optical peak, the X-ray emission
became strongly variable on a timescale shorter than a day and that
the spectrum became soft for a handful of days.  As AG Peg continued
to fade in the optical, the X-ray flux continued to show day to day
variations but the spectrum started to harden. At the time of the
optical re-brightening event the X-ray flux reduced to very low levels
until it started to increase after around 10 days. The degree of
variability reduced and the hardness of the spectrum remained roughly
constant.

Given that the level of absorption did not show an overall
  decrease over the course of the outburst -- as one might expect for
  the clearing of ejecta -- we do not find evidence that the
  day-to-day variability was due to changing levels of absorption from
  a clumpy outflow.  However, we do expect that shocks were generated
in the AG Peg outburst as ejecta interacted with circumbinary material
(see e.g. M\"{u}rset, Wolff \& Jordan 1977, Pan, Ricker \& Taam 2015).
In an extreme case of such interactions, the 2006 outburst of RS Oph
triggered a shock wave that was seen at multiple wavelengths
(e.g., O'Brien et al. 2006). In the 2000--2002 outburst of Z And, a
jet was seen at radio wavelengths (Brocksopp et al. 2004), and it
  is known that jets from symbiotic binaries can produce faint X-rays
  (e.g., Galloway \& Sokoloski 2004; Karovska et al. 2007, 2010,
  Kellogg et al. 2007, Nichols et al. 2007).  Evidence for winds in
  symbiotic binaries, including AG Peg (Nussbaumer et al. 1995), is
  widespread. The optical spectra discussed in \S 2.2 show evidence
  for extended wing features in the Balmer line emission during the
  2015 outburst. For X-ray temperatures $<$1keV, we would expect shock
  speeds slower than 1000 km/s (see eqn 4 of Stute, Luna \& Sokoloski
  2011), consistent with the optical line widths.  Taking a
  Bremsstrahlung cooling time of:

\begin{equation}
 t_{\rm Bremss} = (\frac{T}{2} \times 10^{6} {\rm K})^{1/2} (\frac{n}{3}\times 10^{9} {\rm cm}^{-3}) \sim {\rm days}
\end{equation}

(Frank, King, \& Raine 1992), the rapid X-ray variability suggests
that the shocked gas was probably quite dense. Alternatively, if the
rapid variability was due to adiabatic-expansion cooling of shocked
clumps, the clumps must have been less than $\sim10^{12}$ cm (or
$\sim$1/10 of an AU) in size.  High densities and small clump sizes
are both plausible, so the rapid X-ray variability is consistent with
our hypothesis that the X-ray emission was due to shocks associated
with a (perhaps collimated) eruptive outflow.

Approximately 120 days after the start of the outburst, the optical
flux brightened by $V\sim$0.5 mag in less than a day, after which it
showed a gradual decline over $\sim$40 days followed by a more rapid
decline. A rebrightening event was also seen in the 2000--2002
outburst of Z And which was attributed to a decrease in the
temperature of the optical emitting material due to a slight expansion
of the white dwarf photosphere (Sokoloski et al. 2006). However, the
temperature derived from the He {\sc ii} and H$\beta$ lines (Figure 2)
suggest the temperature did not change significantly over the course
of the rebrightening event. The {\swift} observations of AG Peg show a
suggestion of a $\sim$20 day delay between the initial rise of the
optical and X-ray fluxes at this epoch.

Compared to dwarf nova outbursts, symbiotic binaries show a greater
diversity of X-ray/optical characteristics over an outburst cycle. AG
Dra, whose 1994--1995 outburst was driven by an expansion of the
photosphere of the shell-burning white dwarf, showed a clear drop in
X-ray flux (0.1--2.4 keV) during an optical outburst (Greiner et
al. 1997). On the other hand the symbiotic star V407 Cyg shows an {\it
  increase} in the 0.3--10 keV flux around 20 days after an outburst
in 2010 in which ejecta collided with circumbinary material (Nelson et
al. 2012). Based on the similarity between the optical light curves of
AG Peg and Z And, we suggest that the 2015 outburst of AG Peg was also
triggered by an accretion instability.  Based on the properties of the
X-ray emission, we suggest that enough material was ejected during the
outburst to heat ejecta and/or circumbinary material through shocks.

\subsection{Origin of quasi-steady shell burning in classical symbiotic stars}
\label{shellburning}

M{\"u}rset et al. (1991) and Skopal (2005) modelled UV spectra of a
sample of symbiotic stars obtained using $IUE$ and found white dwarf
temperatures and luminosities around $\sim 10^5$~K and $\sim 10^3$
\lsun respectively --- much higher than expected from accretion alone.
Because nuclear burning releases 40 to 50 times more energy per
nucleon than accretion onto a white dwarf, these high temperatures and
luminosities are naturally explained if shell burning is
present. Given that shell burning is not a common (long-term) feature
of accreting white dwarfs in CVs, why is it so frequently present in
symbiotic stars?

Two main possibilities exist for the origin of quasi-steady shell
burning in normal white dwarf symbiotics.  It could be the result of
accretion at a rate above the theoretical minimum for shell burning
but below the value that leads to expansion of the white dwarf
envelope to giant dimensions (Paczy{\'n}ski \& {\.Z}ytkow 1978; Sion,
Acierno, \& Tomczyk 1979; Paczy{\'n}ski \& Rudak 1980; Fujimoto 1982a;
Nomoto 1982a,b, Iben 1982). One problem with this scenario, however,
is that the range of accretion rates that produce steady burning is
narrow (a factor of a few, Fujimoto 1982b), and it seems unlikely
  that most symbiotic stars accrete at a rate just within this
  range. The other alternative is that the observed quasi-steady shell
  burning is residual burning from a prior nova.

The lower limits that Swift/UVOT placed on the UV flux during the 2015
event attest to the continued presence of shell burning on the surface
of the white dwarf while it experienced classical symbiotic eruptions.  The
$Swift$/UVOT saturates through the UVW2 filter at around 12.6 mag (in
the AB
system\footnote{\url{http://swift.gsfc.nasa.gov/analysis/uvot\_digest/coiloss.html}}),
which corresponds to a flux density at 1928 \AA\hspace{1mm} of $2.75
\times 10^{-13}\,{\rm erg}\,{\rm s}^{-1}\,{\rm cm}^{-2}$\,\AA$^{-1}$.
Multiplying this value by the UVW2 filter width of 657 \AA (Poole et
al. 2008), dereddening it using the $E(B-V) = 0.1$ (Kenyon et
al. 1993), and taking a distance of at least 1~kpc (Boffin et
al. 2014), we find a luminosity between approximately
1600\AA\hspace{1mm} and 2260\AA\hspace{1mm} of at least $4.3 \times
10^{34}\, {\rm erg}\, {\rm s}^{-1}$.  The presence of the 
Raman-scattered O {\sc vi} line (at 6825\AA) throughout the 2015
eruption indicates that the temperature of the hot white dwarf
remained above 110,000~K (M\"{u}rset \& Nussbaumer 1994).  We
therefore multiply the 1600--2260\AA\hspace{1mm} luminosity by a
factor of 5 as a conservative bolometric correction (for a white dwarf
that either hosts shell burning or accretes at a very high rate).
Keeping in mind that both the UVOT flux and the distance are lower
limits, we conclude that the luminosity of either the white dwarf or
the accretion disk was likely well in excess of $\sim 2 \times
10^{35}\, {\rm erg}\, {\rm s}^{-1}$ during the recent outburst.

For the disk to have produced this luminosity it would have required
an accretion rate above $\sim10^{-7}$ \Msun yr$^{-1}$.  It is
theoretically challenging, however, for a 0.6 \Msun white dwarf to
accrete at such a high rate without generating quasi-steady shell
burning (Paczinski \& Zytkow 1978; Nomoto 1982a).  As in other
classical symbiotic stars, such as Z And and AG Dra, the most natural
interpretation is that quasi-steady shell burning remains present on
the surface of the white dwarf.

AG Peg's transition from a slow nova to a classical symbiotic, marked
by classical symbiotic outbursts, suggests that shell burning in at
least some (and perhaps most) symbiotics with strong optical emission
lines is residual burning from prior novae outbursts.  That the
luminosity of the white dwarf in AG~Peg was very low before the 19th
century slow nova (M\"{u}rset \& Nussbaumer 1994) confirms that
material was not burning on the white dwarf before that event.  After
the start of the slow nova, however, the temperature and luminosity
were consistently high enough to indicate shell burning (M\"{u}rset \&
Nussbaumer 1994).  In fact, the slow nova led to a prolonged period of
burning that persisted through the recent active phase.  The lack of
any hard X-ray emission from the accretion-disk boundary layer in 2013
(despite AG~Peg's distance of only about 1~kpc and its exteme UV
brightness; Nu\~{n}ez \& Luna 2013) supports the notion that shell
burning was present through 2013.

As we argued above, the continued UV brightness throughout the recent
period of optical activity implies that shell burning is on-going.
Interestingly, RX~Pup, which has been identified as a symbiotic slow
nova by Mikolajewska et al. (2002), also experienced an optical
rebrightening.  If the steady burning in other symbiotic stars is also
the result of prior novae, it would follow that accretion onto the
white dwarf in these systems does not necessarily proceed at the rate
required for quasi-steady shell burning.  Shell burning arising from
prior novae would also have implications for selection biases in the
sample of known symbiotic stars and the efficiency with which
symbiotic white dwarfs can retain accreted material.

\section{Summary and Conclusions}

The optical outburst of AG Peg in 2015 had a lower amplitude and
  much shorter duration than the outburst observed in the mid 19th
century. The earlier event was very likely driven by a TNR on the
surface of the white dwarf. In contrast, the recent event is similar
to classical symbiotic outbursts in which an accretion
  instability dumps fuel onto an already burning white dwarf.  The
optical light curve of AG Peg is very similar (if taking place over a
shorter timescale) to the 2000--2002 outburst of Z And. The X-ray
observations show strong soft X-ray emission which has large
day-to-day variations 20--30 days after outburst. The X-ray flux
continued to show marked day-to-day variations for another month. This
variability is similar to that seen in some novae such as RS Oph
which is thought to be due to shocks interacting with clumpy material
in the ejecta and emitting X-rays. There is also evidence that
the optical rebrightening event showed X-rays being delayed by
$\sim$20 days compared to the optical, hinting that material had to
clear in order for the X-rays to be seen. Irrespective of the
  exact causes of the 2015 outburst of AG Peg and its X-ray emission,
  it is clear that AG Peg has stopped behaving like a slow nova and
  started behaving like a classical symbiotic star.  With this
  transition, AG Peg provides support for the idea that quasi-steady
  shell burning on the white dwarfs in symbiotic stars is residual
  burning from prior novae.

\section{Acknowledgments}

We warmly thank the {\swift} PI Neil Gehrels and his team for
approving our observations and scheduling them. We acknowledge with
thanks the variable star observations from the AAVSO International
Database contributed by observers worldwide and used in this research.
We also thank Fran{\c c}ois Teyssier for altering us to the many
amateur spectroscopic observations which have been made and we
acknowledge and thank Fran{\c c}ois Teyssier, Umberto Sollecchia, Joan
Guarro Flo, Jacques Montier, Peter Somogyi, Keith Graham and V
Bouttard for use of their spectra. Armagh Observatory is supported by
the Northern Ireland Government through the Dept of Culture, Arts and
Leisure. GJML and NEN acknowledge support from Argentina grant
ANPCYT-PICT 0478/14. GJML and NEN are members of the "Carrera del
Investigador Cientifico (CIC)" of CONICET. We thank the referee
  for a helpful and constructive report.

\appendix 

\section{Tables}

\begin{table*}
\centering
\caption{The observation log for the {\swift} XRT observations. We
  indicate the Observation ID (ObsID); the start of the observations;
  the time since the optical outburst (defined as MJD=57185 = 2015
  June 12); the duration of the XRT exposure and then the observed
  count rate and error (1$\sigma$) in the 0.3--10 keV, 0.3--1 keV and
  1--10 keV energy bands. We have applied a correction to take into
  account that some fraction of the source PSF make have overlapped
  with dead columns on the detector (see text for details).}
\label{xrtlog}
\begin{tabular}{lrrrrcrcrc}
\hline
ObsID & Start Time & Time from    & XRT Exp & 0.3-10 keV & $\pm$ & 0.3-1 keV & $\pm$  & 1--10 keV& $\pm$   \\
      & (MJD)      & Outburst (d) & (s)     & (Ct/s)     &       & (Ct/s)    &        & (Ct/s)   &  \\
\hline
00032906003 & 57201.017 &    16.0 & 8949 &  0.1140  &  0.0045  &  0.0868 &   0.0040 &   0.0197 &   0.0019\\
00032906004 & 57202.128 &    17.1 & 1923 &  0.0951  &  0.0078  &  0.0809 &   0.0072 &   0.0143 &   0.0030\\
00032906005 & 57203.466 &    18.5 & 1627 &  0.0457  &  0.0006  &  0.0321 &   0.0050 &   0.0136 &   0.0033\\
00032906006 & 57204.324 &    19.3 & 1555 &  0.0227  &  0.0044  &  0.0143 &   0.0034 &   0.0080 &   0.0025\\
00032906007 & 57205.667 &    20.7 & 1916 &  0.1063  &  0.0086  &  0.0791 &   0.0074 &   0.0209 &   0.0038\\
00032906008 & 57206.735 &    21.7 & 2018 &  0.0454  &  0.0047  &  0.0392 &   0.0044 &   0.0121 &   0.0025\\
00032906009 & 57207.535 &    22.5 & 1419 &  0.2395  &  0.0140  &  0.2144 &   0.0140 &   0.0267 &   0.0048\\
00032906011 & 57209.784 &    24.8 & 1978 &  0.0297  &  0.0036  &  0.0203 &   0.0030 &   0.0111 &   0.0022\\
00032906012 & 57210.450 &    25.4 & 4864 &  0.1732  &  0.0062  &  0.1620 &   0.0060 &   0.0064 &   0.0012\\
00032906013 & 57211.055 &    26.1 &  903 &  0.3954  &  0.0250  &  0.3803 &   0.0240 &   0.0125 &   0.0044\\
00032906014 & 57212.191 &    27.2 & 1221 &  0.2724  &  0.0180  &  0.2602 &   0.0180 &   0.0069 &   0.0029\\
00032906015 & 57213.258 &    28.3 & 1091 &  0.2776  &  0.0190  &  0.2618 &   0.0190 &   0.0118 &   0.0040\\
00032906016 & 57214.117 &    29.1 & 1622 &  0.0186  &  0.0033  &  0.0168 &   0.0031 &   0.0020 &   0.0011\\
00032906018 & 57217.645 &    32.6 & 1273 &  0.0366  &  0.0062  &  0.0165 &   0.0042 &   0.0221 &   0.0048\\
00032906019 & 57219.042 &    34.0 & 1088 &  0.0406  &  0.0072  &  0.0325 &   0.0064 &   0.0080 &   0.0032\\
00032906020 & 57220.170 &    35.2 & 1715 &  0.2210  &  0.0130  &  0.1995 &   0.0130 &   0.0161 &   0.0036\\
00032906021 & 57221.429 &    36.4 & 1815 &  0.1538  &  0.0096  &  0.1411 &   0.0091 &   0.0095 &   0.0024\\
00032906022 & 57222.095 &    37.1 & 2399 &  0.0136  &  0.0023  &  0.0068 &   0.0017 &   0.0136 &   0.0024\\
00032906023 & 57223.699 &    38.7 & 1602 &  0.0544  &  0.0069  &  0.0476 &   0.0065 &   0.0081 &   0.0027\\
00032906024 & 57224.543 &    39.5 & 1294 &  0.0704  &  0.0089  &  0.0607 &   0.0082 &   0.0081 &   0.0030\\
00032906025 & 57225.627 &    40.6 & 1971 &  0.0191  &  0.0040  &  0.0151 &   0.0035 &   0.0068 &   0.0024\\
00032906026 & 57226.093 &    41.1 & 1822 &  0.1087  &  0.0100  &  0.0935 &   0.0094 &   0.0119 &   0.0033\\
00032906027 & 57227.684 &    42.7 & 1896 &  0.0447  &  0.0055  &  0.0356 &   0.0049 &   0.0089 &   0.0024\\
00032906028 & 57228.748 &    43.7 & 1013 &  0.0564  &  0.0079  &  0.0312 &   0.0058 &   0.0252 &   0.0053\\
00032906029 & 57229.598 &    44.6 & 1976 &  0.0695  &  0.0067  &  0.0434 &   0.0053 &   0.0124 &   0.0028\\
00032906030 & 57230.668 &    45.7 & 1863 &  0.1848  &  0.0120  &  0.1555 &   0.0110 &   0.0152 &   0.0035\\
00032906032 & 57231.883 &    46.9 & 2006 &  0.1975  &  0.0350  &  0.1477 &   0.0310 &   0.0368 &   0.0150\\
00032906033 & 57232.271 &    47.3 & 1361 &  0.0517  &  0.0073  &  0.0340 &   0.0059 &   0.0161 &   0.0041\\
00032906034 & 57233.868 &    48.9 & 1690 &  0.0397  &  0.0062  &  0.0319 &   0.0055 &   0.0125 &   0.0035\\
00032906035 & 57234.468 &    49.5 & 1193 &  0.1532  &  0.0130  &  0.1294 &   0.0120 &   0.0201 &   0.0048\\
00032906036 & 57236.847 &    51.8 & 1951 &  0.2296  &  0.0110  &  0.2037 &   0.0100 &   0.0276 &   0.0037\\
00032906037 & 57237.911 &    52.9 & 1943 &  0.0689  &  0.0068  &  0.0443 &   0.0055 &   0.0233 &   0.0040\\
00032906039 & 57238.588 &    53.6 & 1931 &  0.1519  &  0.0110  &  0.1232 &   0.0098 &   0.0160 &   0.0035\\
00032906041 & 57239.712 &    54.7 & 2016 &  0.0940  &  0.0071  &  0.0704 &   0.0061 &   0.0154 &   0.0029\\
00032906042 & 57240.572 &    55.6 & 1963 &  0.0698  &  0.0071  &  0.0698 &   0.0071 &   0.0207 &   0.0039\\
00032906043 & 57241.636 &    56.6 & 1931 &  0.1068  &  0.0089  &  0.0746 &   0.0074 &   0.0112 &   0.0029\\
00032906044 & 57242.780 &    57.8 & 1968 &  0.0574  &  0.0070  &  0.0455 &   0.0062 &   0.0121 &   0.0032\\
00032906045 & 57243.179 &    58.2 & 1755 &  0.1280  &  0.0110  &  0.0794 &   0.0088 &   0.0131 &   0.0036\\
00032906046 & 57244.498 &    59.5 & 2064 &  0.0435  &  0.0049  &  0.0287 &   0.0039 &   0.0094 &   0.0023\\
00032906047 & 57247.154 &    62.2 &  313 &  0.1657  &  0.0029  &  0.1400 &   0.0270 &   0.0350 &   0.0130\\
00032906048 & 57249.283 &    64.3 & 1900 &  0.0733  &  0.0072  &  0.0590 &   0.0065 &   0.0132 &   0.0031\\
00032906050 & 57253.539 &    68.5 & 2094 &  0.0477  &  0.0055  &  0.0379 &   0.0049 &   0.0103 &   0.0025\\
00032906052 & 57257.089 &    72.1 & 1843 &  0.0492  &  0.0063  &  0.0370 &   0.0054 &   0.0095 &   0.0027\\
00032906053 & 57259.060 &    74.1 & 1707 &  0.0405  &  0.0060  &  0.0333 &   0.0054 &   0.0122 &   0.0033\\
00032906054 & 57266.924 &    81.9 & 1454 &  0.1208  &  0.0110  &  0.1019 &   0.0100 &   0.0174 &   0.0042\\
00032906055 & 57273.188 &    88.2 &  935 &  0.1631  &  0.0130  &  0.1275 &   0.0011 &   0.0217 &   0.0046\\
00032906056 & 57280.477 &    95.5 & 2028 &  0.0894  &  0.0067  &  0.0148 &   0.0027 &   0.0320 &   0.0040\\
00032906057 & 57287.333 &   102.3 & 2006 &  0.1089  &  0.0098  &  0.0768 &   0.0082 &   0.0299 &   0.0051\\
00032906058 & 57302.421 &   117.4 &  948 &  0.1131  &  0.0130  &  0.0890 &   0.0120 &   0.0263 &   0.0064\\
00032906059 & 57306.285 &   121.3 &  970 &  0.0493  &  0.0092  &  0.0576 &   0.0099 &   0.0218 &   0.0061\\
00032906060 & 57309.006 &   124.0 & 1835 &  0.0585  &  0.0066  &  0.0446 &   0.0057 &   0.0124 &   0.0030\\
00032906061 & 57316.531 &   131.5 &  707 &  0.0259  &  0.0074  &  0.0207 &   0.0066 &   0.0090 &   0.0043\\
00032906062 & 57320.926 &   135.9 & 1424 &  0.0533  &  0.0073  &  0.0290 &   0.0054 &   0.0182 &   0.0043\\
00032906063 & 57323.331 &   138.3 &  496 &  0.1048  &  0.0170  &  0.0713 &   0.0140 &   0.0096 &   0.0050\\
00032906064 & 57327.971 &   143.0 & 1492 &  0.1181  &  0.0110  &  0.0957 &   0.0098 &   0.0208 &   0.0045\\
\hline
\end{tabular}            
\end{table*}

\setcounter{table}{0}
\begin{table*}
\centering
\caption{Cont ...}
\label{xrtlog}
\begin{tabular}{lrrrrcrcrc}
\hline
ObsID & Start Time & Time from    & XRT Exp & 0.3-10 keV & $\pm$ & 0.3-1 keV & $\pm$  & 1--10 keV& $\pm$   \\
      & (MJD)      & Outburst (d) & (s)     & (Ct/s)     &       & (Ct/s)    &        & (Ct/s)   &  \\
\hline
00032906065 & 57331.282 &   146.3 & 1387 &  0.1041 &   0.0100 &   0.0620  &  0.0081 &   0.0391 &   0.0064\\
00032906066 & 57335.138 &   150.1 & 2287 &  0.1098 &   0.0081 &   0.0818  &  0.0069 &   0.0289 &   0.0041\\
00032906067 & 57339.794 &   154.8 & 2141 &  0.1072 &   0.0094 &   0.0684  &  0.0069 &   0.0260 &   0.0042\\
00032906068 & 57344.048 &   159.0 & 1094 &  0.1977 &   0.0150 &   0.1380  &  0.0120 &   0.0598 &   0.0081\\
00032906069 & 57349.504 &   164.5 & 2000 &  0.1078 &   0.0089 &   0.0747  &  0.0074 &   0.0291 &   0.0046\\
00032906070 & 57354.884 &   169.9 & 1950 &  0.2048 &   0.0130 &   0.1338  &  0.0100 &   0.0501 &   0.0062\\ 
00032906071 & 57359.806 &   174.9 & 1783 &  0.1803 &   0.0100 &   0.1082  &  0.0078 &   0.0619 &   0.0059\\
00032906072 & 57364.003 &   179.1 & 1893 &  0.1579 &   0.0110 &   0.1262  &  0.0098 &   0.0317 &   0.0049\\
00032906073 & 57369.315 &   184.4 & 1865 &  0.1678 &   0.0120 &   0.1118  &  0.0100 &   0.0436 &   0.0063\\
00032906074 & 57374.169 &   189.4 & 1900 &  0.1663 &   0.0110 &   0.1206  &  0.0094 &   0.0424 &   0.0056\\
00032906075 & 57379.371 &   194.6 & 1908 &  0.1474 &   0.0110 &   0.1114  &  0.0092 &   0.0348 &   0.0052\\ 
00032906076 & 57384.412 &   199.6 & 1034 &  0.1166 &   0.0140 &   0.0711  &  0.0110 &   0.0405 &   0.0081\\ 
00032906077 & 57389.268 &   204.5 & 1401 &  0.0890 &   0.0088 &   0.0716  &  0.0075 &   0.0302 &   0.0052\\ 
00032906078 & 57394.244 &   209.5 & 1968 &  0.1014 &   0.0093 &   0.0766  &  0.0081 &   0.0329 &   0.0053\\ 
00032906079 & 57399.563 &   214.8 & 1778 &  0.1732 &   0.0110 &   0.1193  &  0.0089 &   0.0537 &   0.0060\\
\hline
\end{tabular}            
\end{table*}
            
\end{document}